\input harvmac.tex
\vskip 2in
\Title{\vbox{\baselineskip12pt
\hbox to \hsize{\hfill}
\hbox to \hsize{\hfill IHES/97/P/44}}}
{\vbox{\centerline{Ghost Number Cohomologies and M-theory Quantum States}}}
\centerline{Dimitri Polyakov\footnote{$^\dagger$}
{polyakov@ihes.fr; left for the Cargese Institute May 26th - June 14th}}
\medskip
\centerline{\it Institut des Hautes Etudes Scientifiques}
\centerline{\it 35, Route de Chartres, F-91440 Bures-sur-Yvette, FRANCE}
\vskip .5in
\centerline {\bf Abstract}
We review and develop the formalism of ghost number cohomologies,
outlined in our previous work, to classify the quantum states
of M-theory.We apply this formalism to the matrix formulation
of M-theory to obtain NSR superstring action from
dimensionally reduced matrix model.The BPS condition of the matrix
theory is related to the worldsheet reparametrizational invariance 
in superstring theory, underlining the connection between unbroken
supersymmetries in M-theory and superstring gauge symmetries.

\Date{February 97}
\vfill\eject
\lref\sezgin{E.Sezgin, hep-th/9609086, CTP-TAMU-27-96}
\lref\shenker{D.Friedan,S.Shenker,E.Martinec,Nucl.Phys.{\bf B271}(1986) 93}
\lref\sorokin{I.Bandos, K.Lechner, Nurmagambetov, P.Pasti, 
D.Sorokin, 
M.Tonin, hep-th/9701149 }
\lref\polchino{J.Polchinski,NSF-ITP-95-122,hep-th/9510017}
\lref\townsend{P.Townsend,hep-th/9507048}
\lref\azc{J.A.de Azcarraga, J.P.Gauntlett, J.M.Izquierdo, P.K.Townsend,
 Phys.Rev.{\bf D63} (1989) 2443}
\lref\green{M.B.Green, Phys.Lett.{\bf B223}(1989)157}
\lref\olive{D.Olive, E.Witten, Phys.Lett.78B:97, 1978}
\lref\witten{E.Witten, hep-th/9610234}
\lref\witte{E.Witten, Nucl.Phys.{\bf B463} (1996), p.383}
\lref\myself{D.Polyakov, LANDAU-96-TMP-4, hep-th/9609092,
to appear in Nucl.Phys.{\bf B485}}
\lref\duff{M.J.Duff, K.S.Stelle, Phys.Lett.{\bf {B253}} (1991)113}
\lref\bars{I.Bars, hep-th/9607112}
\lref\grin{M.Green, Phys.Lett. {\bf B223}(1989)} 
\lref\banks{T.Banks, W.Fischler, S.Shenker, L.Susskind, hep-th/9610043}
\lref\seiberg{T.Banks, N.Seiberg, S.Shenker, RU-96-117, hep-th/9612157}
\centerline{\bf 1.Introduction}
\lref\hw{A.Hanany, E.Witten, hep-th/9611230}
\lref\w{E.Witten, hep-th/9610234}
\lref\c{A.Cohn, Non-Commutative Geometry}
\lref\verlinde{R.Dijkgraaf, E.Verlinde, H.Verlinde, hep-th/9704018}
\lref\bs{T.Banks, N.Seiberg, hep-th/9702187}
\lref\bm{T.Banks, L.Motl, hep-th/9703218}
\lref\me{D.Polyakov, hep-th/9703008}
\lref\ian{I.I.Kogan,G.Semenoff,R.Szabo,Mod.Phys.Lett.{\bf A12:183}(1997)}
\lref\nicolai{H.Nicolai, E.Sezgin, Y.Tanii, Nucl.Physics{\bf B305}483 (1989)}
\lref\nicolaiw{B. de Wit, J.Hoppe, H.Nicolai,Nucl.Physics{\bf B305}545 (1988)}
\lref\keh{A.Kehagias, hep-th/9611110}
\lref\ew{E.Witten, Nucl.Phys.{\bf B463}(1996),383}
\lref\nb{N.Berkovits, hep-th 9704109}
Extended superalgebras appear to be a potentially powerful tool
 in describing the dynamics
of branes in M-theory ~\refs{\grin, \azc, \olive, \townsend, \sezgin}
In the recently proposed matrix approach, which supposedly 
describes the M-theory apart from its low-energy
limit ~\refs{\banks, \seiberg}  
the non-perturbative superalgebras play the crucial role.
Non-perturbative superalgebras are superalgebras with p-form central terms,
with p-branes accounting for p-forms.In the matrix theory superalgebra,
 a membrane and a non-covariant fivebrane have been shown to
appear as central terms.M-theory is a strongly coupled limit of 
type IIA  superstring theory ~\refs{\ew}, therefore one may look for the 
string-theoretic origin of the p-forms. 

In the recent paper we have pointed out the connection existing
between picture-changing gauge symmetry in superstring theory
and central terms of non-perturbative superalgebras.
Namely, the certain singular limit of picture-changing transformation
referred to as the ``picture-changing at the infinite-momentum'',plays
the role of ``map'' between non-perturbative and perturbative
superalgebras.

The central terms, generated by this singular version of the
picture-changing, are essentially zero momentum parts of some
peculiar bosonic open string vertex operators. 
These vertices appear to have rather unusual properties -
they are not BRST-trivial, at the same time they do not describe
emissions of any massless particles in perturbative open-string theories.
While their s-matrix elements vanish among elementary string states,
they do interact with Ramond-Ramond charges,i.e.their matrix elements
are non-zero in the presence of D-branes; the examples of such
matrix elements have been computed 
in ~\refs{\me}.These vertex operators may also be interpreted 
as
 `` brane-emitting vertices'',or creation operators for non-perturbative
brane-like states.The example of such an operator is a 5-form
$e^{-3\phi}{\psi_{a_1}...\psi_{a_5}}$ where $\phi$ is a bosonized
superconformal ghost and $\psi_a,a=0,...9$ are 
superstring worldsheet fermions.The crucial feature of
such ``brane-emitting'' vertices is that they all appear to have 
essentially non-zero ghost numbers which cannot be removed by 
picture-changing transformations - and this situation is quite contrary
to the perturbative string theory where it is always possible to 
choose a picture zero for vertex operators.
These observations led us to introduce the notion
of ghost number cohomologies, which we will develop
 here along with 
improving some of the definitions contained in the previous work.
The ghost number cohomologies may be used to classify perturbative
and non-perturbative string states; moreover, we shall argue
that the elements of these cohomologies correspond to the M-theory
quantum states.
Ghost number cohomologies may be used to study the 
relation between matrix theory and strings, which has
been suggested recently ~\refs{\bs,\bm,\verlinde} 
Namely, in this paper we shall demonstrate how one can obtain the 
worldsheet supersymmetric action for the $D=10$ NSR superstring theory
from the matrix theory
by using the correspondence between the M-theory and ghost number 
cohomologies. The analysis will point at 
significantly different roles played by
supersymmetries in the matrix theory and in superstring theory.
The worldsheet reparametrizational invariance in superstring theory
will be related to the BPS condition ~\refs{\banks, \seiberg} in
 matrix theory.
\centerline{\bf Review of Ghost Number Cohomologies}
We start with recalling how the presence of branes modifies the SUSY algebra.
As is now well-known, p-branes  lead to
the appearance of p-form central charges in the superalgebra,
also known as Page charges:
\eqn\lowen{\lbrace {Q_\alpha}, {Q_\beta} \rbrace = \Gamma^m_{\alpha\beta}
P_m + \sum_p \Gamma^{m_1 ... m_p}_{\alpha\beta} Z_{m_1...m_p}}
Strictly speaking , these p-forms are not the central charges since they
may have non-trivial commutation relations with other generators,
as well as with themselves. 

Particularly, for the $D=11$ supermembrane the corresponding
two-form charge is given by:
\eqn\lowen{Z_{m_1m_2}= \int {d^2{\sigma}}\epsilon^{0ik}\partial_i
X_{m_1}\partial_k{X_{m_2}}}
where $X^m$'s are coordinates in the eleven-dimensional
space-time, and the integral is taken over the surface of the membrane.
This charge does not vanish if a membrane
configuration defines non-trivial two-cycles in the space-time.
The presence of this charge is closely related to the fact that
the Wess-Zumino term in the supermembrane action is supersymmetric
only up to total derivative.The integration of the boundary term over
the membrane then gives (2).
 In $D=10$ there exists a surprizing connection between the central terms
in the non-perturbative superalgebra (1) and the singular limit
of perturbative open string gauge symmetry - the picture-changing.
Namely, consider the anticommutator of two supercharges 
$\lbrace{Q_\alpha,Q_\beta}\rbrace$ in ten dimensions, where
the supercharges are given by ~\refs{\shenker}:
\eqn\lowen{Q_\alpha=\oint{{dz}\over{2i\pi}}e^{-{1\over2}\phi}\Sigma_\alpha}
Again, here $\phi$ stands for the bosonized superconformal ghost field
and $\Sigma_\alpha$ is spin operator for matter fields in a space-time.
The O.P.E. between two such spin operators is given by:
\eqn\lowen{\Sigma_\alpha(z)\Sigma_\beta(w)=
{{\epsilon_{\alpha\beta}}\over{(z-w)^{5\over4}}}+\sum_p
{{\Gamma^{a_1...a_5}_{\alpha\beta}\psi_{a_1}...\psi_{a_5}
}\over{(z-w)^{{5\over4}-p}}}+
derivatives}
Then, the straightforward evaluation of the anticommutator gives:
\eqn\grav{\eqalign{\lbrace{Q_\alpha,Q_\beta}\rbrace=
{\Gamma^a_{\alpha\beta}}P_a\cr
P_a=\oint{{dz}\over{2i\pi}}e^{-\phi}\psi_a}}
Here $P_a$ is a momentum operator in $-1$-picture.
In other words, evaluation of the anticommutator with two supercharges
(3) taken in the standard $-{1/2}$-picture gives an usual perturbative
superalgebra (5) with no central terms.
Now, consider another space-time fermionic generator
\eqn\lowen{T_\alpha = \oint{{dz}\over{2i\pi}}e^{-{3\over2}\phi}\Sigma_\alpha}
Contrary to what one may naively suspect, this is $not$ the perturbative
supercharge (3) in another picture, as the straightforward
application of picture-changing operator 
$:\Gamma_1:=:\delta(\gamma)(S_{matter}+S_{ghost}):$ 
(with $S$ being the  worldsheet supercurrent and $\beta,\gamma$
 the superconformal
ghosts)
to (6) gives
zero rather than (3).Nevertheless, as we will show, the generator (6)
$is$ related to the space-time supercharge (3) in a rather subtle way.
  The operator (6) has some interesting  physical properties.
First of all, as we have pointed out some time ago, its integrand generates,
up to picture-changing, the $\kappa$-symmetry transformations
in the Green-Schwarz superstring theory.Also, in the context of extended
$D=10$ space-time superalgebras ~\refs{\grin,\sezgin} it may be understood,
as is easy to check, to be a r.h.s. of the commutator
\eqn\lowen{\lbrack{Q_\alpha,Q_\beta}\rbrack=\Gamma^a_{\alpha\beta}T_\beta}
Evaluating the anticommutator of two $T$'s gives the result:
\eqn\grav{\eqalign{\lbrace{T_\alpha,T_\beta}\rbrace=\oint{{dz}\over{2i\pi}}
\lbrack{1\over2}\Gamma^a_{\alpha\beta}e^{-3\phi}\psi_a({9\over8}\partial
\phi\partial\phi-{3\over2}\partial^2\phi)-{3\over2}\partial{\psi_a}\partial
\phi-{1\over2}\partial^2\psi_a\cr
+\Gamma^{a_1...a_3}_{\alpha\beta}
\partial(e^{\phi}\psi_{a_1}...\psi_{a_3})+\Gamma^{a_1...a_5}_{\alpha\beta}
e^{-3\phi}\psi_{a_1}...\psi_{a_5}\rbrack}}
What is the structure of the r.h.s. of this anticommutator?
As one may check, terms proportional to $\Gamma^a_{\alpha\beta}$
constitute the momentum operator in the $-3$-picture:
applying the picture-changing operator to these terms twice,
one finds,up to unimportant ghost terms:
\eqn\grav{\eqalign{:(\Gamma_1)^2\oint{{dz}\over{2i\pi}}e^{-3\phi}
\psi_a({9\over{16}}\partial\phi\partial\phi-{3\over4}\partial^2\phi)
-{3\over2}\partial\psi_a\partial\phi-{1\over2}\partial^2\psi_a:=\cr
{1\over{16}}\oint{{dz}\over{2i\pi}}e^{-\phi}\psi_a\sim{P_a}}}
The numerical factor of ${1\over{16}}$ is unessential as it can always 
be absorbed by choozing an alternative normalization for $T$.
Therefore we find  that the anticommutator $\lbrace{T_\alpha,T_\beta}\rbrace$
reproduces the non-perturbative superalgebra (1) with the fivebrane
central term proportional to $e^{-3\phi}\psi_{a_1}...\psi_{a_5}$.
At the same time,
the presence of this five-form central term in the non-perturbative
$D=10$ superalgebra is required by the $M$-theory.
There is also a total derivative 3-form term in (8); it is known 
that a threebrane in $D=10,11$ is not fundamental, but rather
just an intersection of two fivebranes. Therefore we interpret
the 3-form of (8) as an intersection term, with the derivative
possibly reflecting the fact of the intersection.
This intersection term ,though not of a fundamental origin,
may be related to monopole dynamics, in the
light of the recently discovered correspondence between
three-dimensional gauge theories and moduli spaces of magnetic 
monopoles ~\refs{\hw} 
We see that $T_\alpha$ may be interpreted as a ``non-perturbative 
supercharge'',generating a superalgebra with branes, with
the map between $T_\alpha$ and $Q_\alpha$ being essentially
the transformation of $S$-duality. Before going further
to explain what this map is and how it is related to the
picture-changing gauge transformation, we note that, although
the two-form term (corresponding to another M-brane - a membrane) is
absent in the anticommutator $\lbrace{T_\alpha,T_\beta}\rbrace$,
it does appear in the anticommutator of $T_\alpha$ with $Q_\beta$:
\eqn\lowen{\lbrace{T_\alpha,Q_\beta}\rbrace=
\Gamma^{a_1a_2}_{\alpha\beta}\oint{{dz}\over{2i\pi}}
e^{-2\phi}\psi_{a_1}\psi_{a_2}+{1\over2}\oint{{dz}\over{2i\pi}}
{\partial}{e^{-2\phi}}\epsilon_{\alpha\beta}}
with the second term apparently related to a $D0$-brane.
In other words,we have obtained a chain of anticommutators
\eqn\lowen{\lbrace{Q_\alpha,Q_\beta}\rbrace\rightarrow
\lbrace{T_\alpha,Q_\beta}\rbrace\rightarrow\lbrace
{T_\alpha,T_\beta}\rbrace}
where  the first anticommutator is 
just a perturbative superalgebra without central terms,the last
one represents the superalgebra with a fivebrane, and
the 

``cross-anticommutator'' $\lbrace{T,Q}\rbrace$ contains a membrane
and a $D0$-brane. 
In order to get a superalgebra which includes both M-branes at once,
one should, of course, simply consider the supercharge
being a sum of $T$ and $Q$:
\eqn\lowen{S_\alpha=T_\alpha+Q_\alpha}
with T being a ``strongly coupled'' part and Q a ``weakly coupled''.
 What precisely relates
$T$ and $Q$? As we have already mentioned, it is not the picture-changing.
Rather,they are connected through quite a peculiar transformation,
of which one may think  as a singular limit of the picture-changing
transform at infinite (or zero) momentum.
Namely, noticing that $T_\alpha$ and $Q_\alpha$  are zero momentum
parts of some fermionic vertex operators,
consider the following vertices:
\eqn\grav{\eqalign{V(k) = \oint{{dz}\over{2i\pi}}
u^\alpha(k)e^{-{1\over2}\phi}\Sigma_\alpha
{e^{ikX}}\cr
W(k,\bar{k})=\oint{{dz}\over{2i\pi}}
v^\alpha(k,\bar{k})e^{-{3\over2}\phi}\Sigma_\alpha
{e^{ikX}}\cr
v_\alpha(k,\bar{k})=(\Gamma{\bar{k}})_{\alpha\beta}u_\beta(k)\cr
(kk)=k^2=({\bar{k}}{\bar{k}})=({\bar{k}})^2=0\cr
(k{\bar{k}})=1}}
Here $k$ is a momentum of the fermionic emission vertex
in the $-{1\over2}$-picture,and $\bar{k}$ is an auxiliary momentum,
analogous to the one present in the dilaton vertex operator;
$u_\alpha(k)$ is some constant space-time spinor, satisfying
the on-shell Dirac equation.
The obvious difficulty with our definition of the vertex operator
$W(k,\bar{k})$ is that it is not BRST-invariant because of our
choice of its polarization spinor $v_\alpha(k,\bar{k})$.
In the limit $\bar{k}\rightarrow\infty$ and
$k\rightarrow{0}$, however, its
BRST-invariance is restored and the operation of picture-changing
is again well-defined.
Applying the picture-changing operator now gives:
\eqn\grav{\eqalign{lim_{{\bar{k}}\rightarrow\infty}:\Gamma_1W(k,\bar{k}):=
i(\Gamma{\bar{k}})(\Gamma{k})_{\alpha\gamma}\oint{{dz}\over{2i\pi}}
e^{-{1\over2}\phi}\Sigma_\gamma{e^{ikX}}\cr=
2iu_\alpha(k)\oint{{dz}\over{2i\pi}}e^{-{1\over2}\phi}\Sigma_\alpha
e^{ikX}=2iV(k)}}
where we have used the identity
$(\Gamma\bar{k})(\Gamma{k})+(\Gamma{k})\Gamma\bar{k})=2(k\bar{k})=2$
and the on-shell condition for the spinor $u_\alpha(k)$.    

Therefore the relation between $T_\alpha$  and $Q_\alpha$
is given by:
\eqn\lowen{N_{\alpha\beta}T_\beta=
{((\Gamma{\bar{k}})(\Gamma{k}))_{\alpha\beta}}
Q_\beta}
where the S-duality generator $N_{\alpha\beta}$ is defined as
\eqn\lowen{N_{\alpha\beta}=lim_{\bar{k}\rightarrow\infty}
\lbrack(\Gamma{\bar{k}})_{\alpha\beta}:\Gamma_1:e^{ikX}\rbrack}
Another role of this generator is that it ``improves'' the
operation of picture-changing, in general not well-defined at $k=0$.
It should be emphasized that the operation $N_{\alpha\beta}$ is
only defined for those zero momentum fermionic vertices belonging
to the kernel of $\Gamma_1$ .For the operators not
belonging to  $ker(\Gamma_1)$, $N_{\alpha\beta}$ should be
replaced by ordinary picture-changing.
Let us now analyze the central $p$-form terms appearing in the 
non-perturbative superalgebra (8). The 5-form
$e^{-3\phi}\psi_{a_1}...\psi_{a_5}$ appears to be a zero momentum part
of a  rather peculiar vertex operator, which, while not being BRST trivial,
does not appear to describe an emission of any massless particle in
perturbative string theory.Its
S-matrix elements
vanish among elementary string states, but are non-zero in the presence
of D-branes due to the interaction of this vertex with Ramond-Ramond
charges.This property prompts us to interpret this 5-form as
a brane-emitting vertex (versus particle-emitting vertices
in perturbative string theory).
Another crucial property of this vertex operator is that it
has no analogue in the picture zero (although it does have an analogue in
the $+1$-picture), in other words its nonzero ghost number appears
to be its indispensable feature - and again this is quite contrary to the
properties of perturbative string vertices for which the representation in
the picture of ghost number zero always exists. 
 The above considerations lead us to introduce the following
notion of ghost number cohomologies.
Let $\lbrace{V_n}\rbrace$ 
 be a set of physical states (vertex operators),perturbative
or non-perturbative, having a ghost number $n\leq{0}$
For $n<0$, let us further define the subset 
$\lbrace{\tilde{V_n}}\rbrace\subset\lbrace{V_n}\rbrace$
of the operators of ghost number $n$ for which there exists a picture-changing
transformation relating them to vertices of some ghost number
$m$, $n<m\leq{0}$,i.e.$\exists{V_m}\in\lbrace{V_m}\rbrace,n<m\leq{0}:$
\eqn\lowen{V_m=:(\Gamma_1)^{m-n}:V_{n}\in\lbrace{\tilde{V_n}}\rbrace}
By definition, we put $\lbrace\tilde{V_0}\rbrace=\emptyset$
The ghost number $n\leq{0}$ cohomologies $\lbrack{H_n}\rbrack$
are then defined as
\eqn\lowen{\lbrack{H_n}\rbrack={{\lbrace{V_n}\rbrace}\over
{\lbrace{\tilde{V_n}}\rbrace}}}  
All the states not belonging to any of $\lbrack{H_{n}}\rbrack$ of
$n\neq{0}$ are by definition set to belong to $H_{0}$
The fivebrane central 
term $\oint{{dz}\over{2i\pi}}e^{-3\phi}\psi_{a_1}...\psi_{a_5}$
is then the element of $\lbrack{H_{-3}}\rbrack$;
the membrane $\oint{{dz}\over{2i\pi}}e^{-2\phi}\psi_{a_1}\psi_{a_2}$
of the anticommutator $\lbrace{T_\alpha,Q_\beta}\rbrace$
belongs to $\lbrack{H_{-2}}\rbrack$.
However, for instance the operator $\oint{{dz}\over{2i\pi}}e^{-\phi}\psi_a$
does not belong to to $\lbrack{H_{-1}}\rbrack$, as the picture-changing
operator $\Gamma_1$ transforms it into $\oint{{dz}\over{2i\pi}}\partial{X_a}$
of $\lbrack{H_0}\rbrack$.
The cohomology $\lbrack{H_{-4}}\rbrack$ contains the non-dynamical
state defined by
$\oint{{dz}\over{2i\pi}}e^{-4\phi}\psi_{a_1}...\psi_{a_{10}}$ 
which arguably may be attributed
to cosmological constant.
Other ghost number cohomologies seem to be redundant; our conclusion is that
 $\lbrack{H_{-4}}\rbrack,\lbrack{H_{-3}}\rbrack,\lbrack{H_{-2}}\rbrack$,
$\lbrack{H_0}\rbrack$ form a basis for the quantum states of $M$-theory.
 All the elementary string states belong
to $\lbrack{H_0}\rbrack$, while the non-perturbative physics is
hidden in the cohomologies of non-zero ghost numbers.
Thus, the cohomologies $\lbrack{H_{-2}}\rbrack,\lbrack{H_{-3}}\rbrack$
contain D0-branes and M-brane states (including intersecting branes),
 and $\lbrack{H_{-4}}\rbrack$
accounts for the cosmological constant.
Let us present this schematically:
\eqn\grav{\eqalign{\lbrack{H_0}\rbrack\rightarrow
{particles}\cr
\lbrack{H_{-2}}\rbrack\rightarrow{membranes+D0-branes}\cr
\lbrack{H_{-3}}\rbrack\rightarrow{5-branes}\cr
\lbrack{H_{-4}}\rbrack\rightarrow{cosm.constant(?)}}}

This classifies the M-theory quantum
states in  the formalism of ghost number cohomologies.
Dualities are contained in maps between these cohomologies;
the example of such a map is the $N$-operator (16),
which relates  elementary and non-perturbative superalgebras.

In the following section we will
 explore the application of this formalism to 
 the matrix theory.

\centerline{\bf Application to Matrix Theory}

The action of the matrix model
is given by the $N\times{N}$ matrix ~\refs{\banks}:
\eqn\lowen{{1\over{2g}}\int{dt}({({D_0}X^a)^2}+\theta^\alpha
{D_0}\theta^\alpha+{R\over4}{{\lbrack{X^a,X^b}\rbrack}^2}+
{{iR}\over2}\lbrack{\theta^\beta,\lbrack{X^a},\theta^\alpha\rbrack\rbrack
{\Gamma^a_{\alpha\beta}}})}
where $D_0=\partial_0-i\lbrack{A_0},\rbrack$ and
$\theta^\alpha$($\alpha=1,...16$),$X^a$,$a=1,...10$ and $A_0$
are hermitian $N\times{N}$ matrices.
The dynamical SUSY transformations are given by:
\eqn\grav{\eqalign{\delta{X^a}=-2\epsilon^\alpha\theta^\beta
{\Gamma^a_{\alpha\beta}},\cr
\delta\theta_\alpha={1\over2}(D_0{X_a}\Gamma^a_{\alpha\beta}\epsilon^\beta+
\lbrack{X_a,X_b}\rbrack\Gamma^{ab}_{\alpha\beta}\epsilon^\beta)\cr
\delta{A_0}=-2\epsilon^\alpha{\theta_\alpha}}} 
and the trivial kinematic SUSY is given by
\eqn\lowen{\delta{\theta^\alpha}=\epsilon^\alpha,\delta{X^a}=\delta{A_0}=0}
Now, let us reduce the Lagrangian (20) to $D=0$,i.e. to 
a zero-dimensional $N\times{N}$ matrix model.Such a reduction implies
choosing the gauge $A_0=0$ and dropping the kinetic terms in (20).
The dimensionally reduced Lagrangian is given by
\eqn\lowen{{{1}\over{2g}}tr(\sum_{a<b}{(\lbrack{X_a,X_b}\rbrack)^2}-
\sum_a\lbrack
{\theta^\alpha\Gamma^a_{\alpha\beta}\lbrack{\theta^\beta,X_a}\rbrack}\rbrack)}
The superalgebra corresponding to the supersymmetry transformations
(21),(22) contains the following 2-form central charge ~\refs{\seiberg}:
\eqn\lowen{Z_{ab}= {i\over2}Tr\lbrack{X_a, X_b}\rbrack}
The energy of the membrane state is proportional to 
to the square of the membrane charge:
\eqn\lowen{E\sim(Z_{ab})^2 = {-{1\over4}}(Tr\lbrack{X_a,X_b}\rbrack)^2}
In the matrix-superstring relation this 
expression for the membrane energy should correspond
to the stress-energy tensor in superstring theory
(here the  appropriate
dimensional reduction from $D=11$ to $D=10$ is implied, of course,
to produce a string out of a membrane)
 To construct a superstring stress-energy tensor from (25) we therefore need
a ``glossary'' which relates the matrix variables of (20),(23) to the
 superstring variables in the NSR formalism, under the reduction to $D=10$.
So what is the counterpart of the commutator
 $\lbrack{X_a,X_b}\rbrack$ in the strongly coupled superstring theory?
Since $\lbrack{X_a,X_b}\rbrack$ is the membrane topological charge
in matrix theory, it corresponds, in the formalism of 
ghost number cohomologies, to the membrane of $\lbrack{H_{-2}}\rbrack$,
i.e. we require
\eqn\lowen{\lbrack{X_a,X_b}\rbrack\rightarrow{e^{-2\phi}\psi_a\psi_b}}
The  drawback of such an ansatz is that it does 
not seem to preserve the identity satisfied by the matrices
$X_a$ due to the BPS property ~\refs{\seiberg}:
\eqn\lowen{\lbrack{X_a,X_b}\rbrack={i\over2}\epsilon_{abcd}
\lbrack{X_c,X_d}\rbrack}
Nevertheless, while the right-hand side of (26) does not satisfy the BPS
identity (27) in a straightforward way, we will see later that
in superstring theory the analogue of (27) does hold , corresponding
to  the stress-energy tensor vanishing 
(which is equivalent, in turn, to the worldsheet
reparametrizational invariance
condition)
The Sugawara stress-energy tensor corresponding to (25) under such a
matrix-superstring correspondence is then given by:
\eqn\grav{\eqalign{T^{(-4)}(z) =  
:e^{-2\phi}\psi_a\psi_b e^{-2\phi}\psi_a\psi_b:(z)=\cr
e^{-4\phi}\lbrace({1\over2}\psi_a{\partial^2}\psi_a\psi_b{\partial^2}\psi_b
+{1\over3}\psi_a{\partial^3}\psi_a\psi_b\partial\psi_b-{1\over{15}}
\psi\partial^5\psi+{4\over{45}}P^{(6)}(-2\phi))+\cr
(\psi_a{\partial}\psi_a\psi_b\partial\psi_b-{1\over3}\psi\partial^3\psi)
(\partial^2\phi-2\partial\phi\partial\phi)+
(\psi_a\partial^2\psi_a\psi_b\partial\psi_b\partial\phi
-{2\over3}\psi\partial^4\psi)
\rbrace}}
where $(-4)$ stands for the total ghost number $-4$ of this tensor.

At first glance this expression does not seem at all to resemble  the
NSR stress-energy tensor.To understand its relevance to 
superstring theory, it is crucial to point out the role played 
by the fermionic part of the matrix theory Lagrangian (23),which is 
given by the double commutator 
$\Gamma^a_{\alpha\beta}\lbrack\theta^\alpha,\lbrack\theta^\beta,X_a\rbrack
\rbrack$.
It has been shown \refs{\ian} that the integration over fermionic 
variables $\theta^\alpha$ in the partition function with the
static part of the action (3) yields the determinant 
$Pfaff({i\over{2g}}f_{ijk}{\sum_a}\gamma^a_{\alpha\beta}X_a^k)$ 
This Pfaffian has been shown to identify the Nicolai map ~\refs{\nicolai,
\nicolaiw}
for the bosonic potential $Tr\lbrack{X_a,X_b}\rbrack$.
That is, the Nicolai map : $W^c=\gamma^{c}_{k{\dot{k}}}\lbrack{X_k,
X_{\dot{k}}}\rbrack$
defines the new variable $W^c$ in terms of the membrane charge
$\lbrack{X,X}\rbrack$,and the Jacobian of this map precisely 
cancels the Pfaffian coming out of the integration over fermions.
The partition function then reduces to the 
finite-dimensional Gaussian integral over
$W$.
Next, we observe the following important connection between
Nicolai map and picture-changing gauge transformation.Namely,
consider the gauge-fixed $D=10$
supersymmetric fivebrane Lagrangian in a light-cone gauge:
\refs{\nicolai}
\eqn\grav{\eqalign{I={{T_5}\over2}DX^a{DX^a} - det(\partial_r{X^a}
\partial_s{X^a})+i{\bar\theta}D\theta\cr+{i\over4}\epsilon^{rstuv}
\partial_r{X^{a_1}}\partial_s{X^{a_2}}\partial_t{X^{a_3}}\partial_u
{X^{a_4}}\bar\theta\gamma_{a_1a_2a_3a_4}\partial_v\theta}}
For this fivebrane action the equilibrating Nicolai map is given by
~\refs{\keh}:
\eqn\lowen{|\eta_a|= {{dX_a}\over{d\tau}}+{1\over{5!}}
\epsilon^{rstuv}\epsilon_{a{a_1}...a_5}\partial_r{X^{a_1}}...
\partial_v{X^{a_5}}}
 The quintic-like Wess-Zumino term in the fivebrane
action (29) is understood to give rise to the five-form central
term in the superalgebra (1), with $p=5$.
After having performed the Nicolai transform (30) we obtain
an essentially quadratic action in terms of new variables $\eta$,
without the Wess-Zumino term - and accordingly, the superalgebra
in the Nicolai-transformed theory should no longer include the five-form
central term.In other words, the  Nicolai transform
 maps the superalgebra
with the central 5-form term to the one
without central terms - but this is
exactly what the $N$-operator (16) also does.
This motivates
our conjecture that the string-theoretic counterpart 
of the fermionic term in (23) must be the
picture-changing operator $:\Gamma_1: = :e^\phi(S_{matter}+S_{ghost}):$
of conformal dimension zero.
Beside that, the following heuristic argument may be given.
 Suppose that the counterpart of the fermionic matrix variable
$\theta^\alpha$ is the Green-Schwarz fermion $\theta^\alpha(z)$.
Then it is related to the NSR variables through
\eqn\lowen{\theta^\alpha = e^{\phi\over2}\Sigma^\alpha + {ghosts}}, 
where $\Sigma^\alpha$ is a
spin operator for matter fields. Then, 
the GSO projected string-theoretic counterpart of the fermionic term,
having conformal dimension zero is
\eqn\grav{\eqalign{
\Gamma^a_{\alpha\beta}\lbrack\theta^\alpha,\lbrack\theta^\beta,X_a\rbrack
\rbrack\rightarrow\Gamma^a_{\alpha\beta}:e^{\phi\over2}\Sigma^\alpha(z)
e^{\phi\over2}\Sigma^\beta(w){X_a}:=e^{\phi}Tr(\Gamma^a\Gamma^m)\psi_m\partial
{X_a}\cr=8e^\phi\psi_a\partial{X_a}+ghosts \sim{\Gamma_1}}}

i.e. it is proportional to the picture-changing
 operator $\Gamma_1$.
The superstring partition function corresponding to 
the reduced matrix theory action (23) is then given by
\eqn\lowen{Z=\int{D}\lbrack{X}\rbrack{D}\lbrack{X}\rbrack{D}\lbrack{ghosts}
\rbrack{e^{\Gamma_1}}e^{-S^{(-4)}}}
where $S^{(-4)}$ is the action corresponding to the Sugawara tensor 
 $T^{(-4)}$ of (28).
Then, the partition function may be written as:
\eqn\grav{\eqalign{Z= 
\int{D}\lbrack{X}\rbrack{D}\lbrack{X}\rbrack{D}\lbrack{ghosts}\rbrack
\lbrace\sum_m{{:(\Gamma_1)^m:}\over{m!}}\sum_n{{(-S^{(-4)})}\over{n!}}
\rbrace\cr=\int{D}\lbrack{X}\rbrack{D}\lbrack{X}\rbrack{D}\lbrack{ghosts}
\rbrack\sum_n{{({-:(\Gamma_1)^4:S^{(-4)}})^n}\over{n!}}\cr=
\int{D}\lbrack{X}\rbrack{D}\lbrack{X}\rbrack{D}\lbrack{ghosts}\rbrack
e^{-S^{(0)}}}}
where $S^{(0)}=:(\Gamma_1)^4{S^{(-4)}}:$
due to ghost number conservation.
It is easy to check now that the stress-energy tensor
$T^{(0)}$ corresponding to the ``effective'' action $S^{(0)}$ is
given by 
\eqn\lowen{T^{(0)}=:(\Gamma_1)^{4}{T^{(-4)}}:}
Indeed, since $T_{ik}={{\delta{S}}\over{\delta{\gamma^{ik}}}}$,
(35) simply follows from the fact that variation of the picture-changing
operator with respect to the worldsheet metric vanishes. 
Applying the picture-changing operator $\Gamma_1$ four times
to $T^{(-4)}$ we obtain
\eqn\grav{\eqalign{T^{(0)}=:(\Gamma_1)^4T^{(-4)}:=
{1\over2}(\partial{X^m}\partial{X_m}+\partial\psi^m\psi_m+
\partial\sigma\partial\sigma+3\partial^2\sigma\cr-\partial\phi
\partial\phi-2\partial^2\phi+\partial\chi\partial\chi)}}
where $\sigma$ is bosonized fermionic  ghost:
$c=e^{\sigma},b=e^{-\sigma}$
Thus $T^{(0)}$  is exactly  the expression for
the full matter$+$ghost stress-energy tensor of the NSR superstring theory in
ten dimensions and therefore $S^{(0)}$ is the NSR superstring  in the
superconformal gauge.
We see that the supersymmetry in the matrix theory and the worldsheet
supersymmetry in the NSR superstring theory appear on essentially
different grounds.That is, naively one may expect the supersymmetries
of the matrix theory (21),(22) to translate into the 
worldsheet supersymmetry of the NSR theory in this matrix-superstring
correspondence through ghost number cohomologies.
However, it appears that the
entire worldsheet supersymmetry of superstring theory is 
contained in just the bosonic part of the matrix theory Lagrangian,
given by the term $\sim(\lbrack{X_a,X_b}\rbrack)^2$ as this
term accounts for the Sugawara tensor (28) , of which the
NSR stress-energy tensor (36) is obtained by the 
four-fold application of the picture-changing
operator $\Gamma_1$.Now, the fermionic part of the matrix
theory action (23) is exactly the one which gives us the
picture-changing transformation we need.
  Let us now address the question of compatibility of the matrix theory
BPS condition (27) and the ansatz (26) dictated by ghost number cohomology
arguments.At first glance, there seems to be a disagreement
between (26) and (27), but this disagreement may be resolved due
to what we find to be an intriguing fact in the matrix-superstring
correspondence.
Let us multibly both the left-hand and the right-hand sides of (27)
by $\lbrack{X_a,X_b}\rbrack$, substitute the ansatz (26) and
apply the four-fold picture-changing $({\Gamma_1})^4$ to both
the l.h.s. and the r.h.s. of the identity obtained.
Then, as we have already shown, the left-hand side
 becomes the worldsheet NSR
stress-energy tensor $T(z)$. As to the 
right-hand side, due to the ansatz (26)
it becomes:
\eqn\grav{\eqalign{-{1\over4}\epsilon_{abcd}\lbrack{X_a,X_b}\rbrack
\lbrack{X_c,X_d}\rbrack=-{1\over{4}}\epsilon_{abcd}e^{-4\phi}
\lbrace\partial^4({\psi_a\psi_b})\psi_c\psi_d
-{1\over3}\partial\phi\partial^3(\psi_a\psi_b)\psi_c\psi_d\cr
+(\partial\phi\partial\phi-{1\over2}\partial^2\phi)\partial^2(\psi_a\psi_b)
\psi_c\psi_d-({4\over3}\partial\phi\partial\phi\partial\phi-
{4\over3}\partial^2\phi
\partial\phi+{1\over3}\partial^3\phi)\partial(\psi_a\psi_b)\psi_c\psi_d\cr+
{1\over{24}}(-2\partial^4\phi+8(\partial^2\phi\partial^2\phi+\partial^3\phi
\partial\phi)-24\partial^2\phi\partial\phi\partial\phi+16\partial\phi\partial
\phi\partial\phi\partial\phi)\psi_a\psi_b\psi_c\psi_d\rbrace}}
Applying the operator $:(\Gamma_1)^4:$ to the right-hand side of
(37) one finds that the picture zero counterpart of (37) vanishes.
Therefore, we find that the BPS relation (27) in the matrix theory translates
into the condition $T=0$ in the superstring theory,i.e. the condition
of reparametrizational invariance on the worldsheet.
This result points at the connection between
the unbroken supersymmetries of the matrix model and worldsheet 
gauge symmetries in superstring theory.

\centerline{\bf Conclusion}

In this paper, only those $p$-form fields of $n\neq{0}$ ghost
number cohomologies  corresponding to ground states of branes have been
considered.The next logical step would be to extend this analysis 
to involve other p-brane modes of $\lbrack{H_n}\rbrack$ $n\neq{0}$.
The method of ghost number cohomologies  appears to be helpful
in analyzing the non-perturbative spectrum of superstring theory and
M-theory.The non-perturbative physics is hidden in cohomologies
of non-zero negative numbers, while $\lbrack{H_0}\rbrack$ contains
the elementary states of superstring theory. 
The conformal field theory may be used to 
analyze the scattering amplitudes of branes.
Applying this method to the matrix theory emphasizes the connection
between matrix M-theory and superstrings. Some novel features 
arise -
such as the worldsheet reparametrizational invariance 
corresponding to  the matrix theory BPS condition  and 
the supersymmetry of the dimensionally reduced matrix model 
being the analogue of picture-changing in superstring theory.
In general,
dualities may be understood as maps between ghost number cohomologies.
We hope that studying the structure of ghost number cohomologies
may prove helpful to understand the interplay between M-theory and
non-commutative geometry, suggested originally in ~\refs{\banks}
We hope to elaborate on that in 
our future works.
Very roughly, one may consider ghost number cohomologies as foliated
spaces with picture-changing and $N$-operators
playing the role of foliations.
P-branes then may be interpreted as leaves of integrable foliations.
Given the string-theoretic origin of the $p$-form terms in (1)
the proper question is where the eleventh dimension comes from.
Recently the discussion in ~\refs{\nb} has pointed at the role
that the twistor-like superstring variable plays
in ``building the bridge'' 
to $D=11$.
If the cohomologies of higher
ghost numbers are to adequately describe the non-perturbative physics of 
 M-theory, and the fivebrane dynamics in particular,
their structure shall somehow
involve  the 2-form non-Lagrangian field
 propagating on the fivebrane worldvolume, which 
partition function has been determined in ~\refs{\w}
At present, finding such a correspondence is an open question. 

\centerline{\bf Acknowledgements}

The present work is supported by the European Post-Doctoral Institute
(EDDI).The author acknowledges the hospitality of the Institut
des Hautes Etudes Scientifiques (IHES).I'm grateful to 
A.Cohn and T.Damour for discussions.
\listrefs 
\end